\documentclass[prl,aps,twocolumn,showpacs,floatfix]{revtex4}

\usepackage{mathrsfs}
\usepackage{color}
\usepackage{graphicx}
\usepackage{hyperref}
\usepackage{hypernat}
\usepackage[latin1]{inputenc}
\usepackage{pstricks}
\usepackage{amsmath,amssymb}
\usepackage{version}
\usepackage[outdir=./]{epstopdf}
\DeclareGraphicsExtensions{.eps,.ps}
\usepackage{ulem}
\usepackage{float}
\usepackage{booktabs}
\usepackage{subfigure}
\usepackage{dcolumn}
\usepackage{bm}

\begin{document}

\preprint{APS/123-QED}
\title{Spin-1 Bosons in the Presence of Spin-orbit Coupling}
\author{Lin Xin$^{1}$}
\email{lxin9@gatech.edu}
\affiliation{
 $^1$ School of Physics, Georgia Institute of Technology, Atlanta, GA 30332, U.S.A\\
}

\date{\today}

\begin{abstract}
In this paper, I'm going to talk about the theoretical and experimental progress in studying spin-orbit coupled spin-1 bosons.  Realization of spin-orbit coupled quantum gases opens a new avenue in cold atom physics. In particular, the interplay between spin-orbit coupling and inter-atomic interaction leads to many intriguing phenomena. Moreover, the non-zero momentum of ground states can be controlled by external fields, which allows for good quantum control. 
\end{abstract}

\maketitle

Spin-orbit (SO) coupling is named because it's the interaction between a quantum particle's spin and its angular momentum. It comes from the conversion between electric field to magnetic field when changing into the moving frame of particles. It is important because when dealing with neutral atoms, the interaction between spin and magnetic field dominates. A key point of this paper is to emphasize that a nearly isotropic SO coupling will dramatically enhance the effects of inter-particle interactions.

There are two types of interaction that have been studied extensively in condensed matter - Rashba ($\sigma_{x}k_{y}-\sigma_{y}k_{x}$) \cite{Rashba} and Dresselhaus ($\sigma_{x}k_{y}+\sigma_{y}k_{x}$) \cite{Dresselhaus} SO coupling.These two coupling have explained lots of phenomena for bulk materials.  The Rashba SO coupling is a momentum-dependent splitting of spin bands in two-dimensional condensed matter systems (heterostructures and surface states) similar to the splitting of particles and anti-particles in the Dirac Hamiltonian. It can be derived in the framework of the $k\cdot p$ perturbation theory or from the point of view of a tight binding approximation. The Dresselhaus SO coupling can be derived in a similar way. The most general expression of Rashba SO coupling is presented below:

\begin{equation}
H_{SO}=2\alpha(\vec{\sigma} \times \vec{k})\cdot \vec{v}.
\end{equation}

Here $v$ is the unit vector perpendicular to the surface, $k$ is the momentum of particle and $\sigma$ is the vector of Pauli matrices.

There is a naive derivation of Rashba Hamiltonian for particles moving in static electric fields. For example, $E=E_{0}\hat{z}$. When we use the frame moving with atoms, where $v=\frac{\hbar}{m}(k_{x},k_{y},k_{z})$. By applying the Maxwell function in a moving frame.
\begin{align}
\vec{E'}=\vec{E}+\vec{v}\times \vec{B}\\
\vec{B'}=\vec{B}-\frac{\vec{v}\times \vec{E}}{c^{2}}
\end{align}

In the new frame, we have
\begin{align}
\vec{E'}=E_{0}\hat{z}\\
\vec{B'}=\frac{E_{0}}{mc^{2}}(-k_{y},k_{x},0).
\end{align}

Thus, the Hamiltonian has a term between the spin and interaction with the magnetic field caused by momentum.
\begin{align}
H_{SO}=-\vec{\mu}\cdot \vec{B_{SO}}(k)=-\frac{g_{s}\mu_{B}S}{\hbar}\cdot\frac{E_{0}}{mc^{2}}(-k_{y},k_{x},0)\\
=-\frac{g_{s}\mu_{B}E_{0}}{\hbar mc^{2}}(\sigma_{x}k_{y}-\sigma_{y}k_{x})
\end{align}
where $g_{s}$ is Land$\acute{e}$ g factor and $\mu_{B}$ is Bohr magneton.

As a result the total Hamiltonian for a single boson only under the presence of electric field is : 
\begin{equation}
H=\frac{\hbar^{2}}{2m}(\vec{k}^{2}+2\alpha (\vec{\sigma} \times \vec{k})\cdot \hat{z}).
\end{equation}
Here all the constants are combined in $\alpha$. In general situation with arbitrary $\vec{E}$ point in $\hat{n}$ direction and with atomic units $\hbar \to 1, m_{e} \to 1$
\begin{equation}
H=\frac{1}{2m}(\vec{k}^{2}+2\alpha (\vec{\sigma} \times \vec{k})\cdot \hat{n}).
\end{equation}

The electric field given by a nucleus in the rest frame of the electron is:
\begin{equation}
\bm{E}=|\frac{E}{r}|\bm{r}.
\end{equation}
Additionally, since $|E|=\frac{\partial V}{\partial r}$
\begin{equation}
\bm{B}=\frac{1}{m_{e}c^{2}}\frac{1}{r}\bm{r}\times\bm{p}\frac{\partial V}{\partial r}=\frac{1}{m_{e}c^{2}}\frac{1}{r}\frac{\partial V}{\partial r}\bm{L}.
\end{equation}
As a result,
\begin{align}
H&=-\mu\cdot B\\
&=\frac{1}{m_{e}c^{2}}\frac{1}{r}\frac{\partial V}{\partial r}\bm{L}\cdot -g_{s}\mu_{B}\frac{\bm{S}}{\hbar}=-\frac{g_{s}\mu_{B}}{m_{e}c^{2}r\hbar}\frac{\partial V}{\partial r}\bm{L}\cdot \bm{S}.
\end{align}
The Larmor interaction energy is in the form of $L\cdot S$ coupling, that's the reason why it's called spin-orbital coupling.

In the second quantized picture, the Hamiltonian is
\begin{equation}
H=\int d^{2} r\Psi^{\dagger}\frac{1}{2m}(\vec{k}^{2}+2\alpha(\vec{\sigma} \times \vec{k})\cdot \vec{n})\Psi
\label{2nd quantized H}
\end{equation}

\begin{figure}
\includegraphics[width=0.3\textwidth]{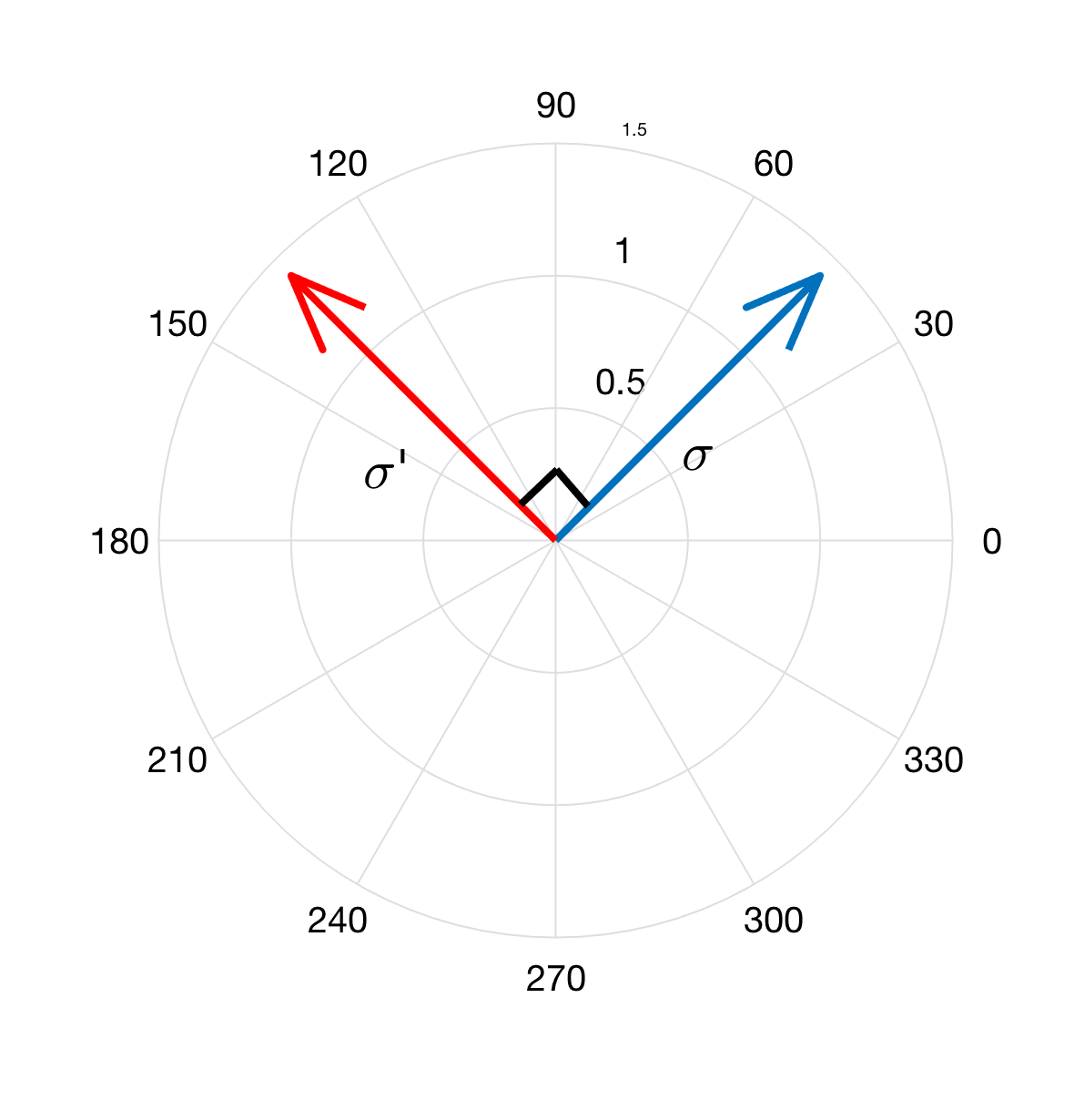}
\caption{(color online) Pseudo-spin rotation rotates the real spin by $90^{o}$ anticlockwise.}
\label{pseudo}
\end{figure}

 In many other literature \cite{review} people use pseudo-spin rotation $\sigma_{x}'=-\sigma_{y}$ and $\sigma_{y}'=\sigma_{x}$ to simplify problem. This rotation will change the cross product between $\sigma$ and $k$ into a dot product, as shown in Fig. \ref{pseudo}. Here, I follow this conversion. The Hamiltonian will be 
 
\begin{equation}
H=\int d^{2} r\Psi^{\dagger}\frac{1}{2m}(\vec{k}^{2}+2\alpha(\vec{\sigma} \cdot \vec{k}))\Psi
\label{2nd quantized H}
\end{equation}

In the first part of this paper, what I will focus on is the ground state for a two dimensional (2D) Bose-Einstein Condensates (BEC) and study different phases of them.

\subsection{Numerical Method}
In order to solve the ground state for general Hamiltonian, using a numerical method is a very popular way to attack problem, given the advance of computer performance nowadays. The method used here is called the Imaginary Time Evolution method \cite{Imaginary}. The physics behind that is straightforward. Decomposing the initial wave function into the eigenbasis, where $\phi_{0}$ is the ground state with lowest energy and $\phi_{n}$ is the $n^{th}$ excited state, I have
\begin{equation}
|\psi>=c_{0}|\phi_{0}>+c_{1}|\phi_{1}>+c_{2}|\phi_{2}>+\cdots
\end{equation}
The time evolution of the wave function is
\begin{equation}
|\psi(t)>=c_{0}e^{-i\frac{E_{0}}{\hbar}t}|\phi_{0}>+c_{1}e^{-i\frac{E_{1}}{\hbar}t}|\phi_{1}>+\cdots
\end{equation}
If I change the time $t$ into imaginary time $-it$, the time evolution will become
\begin{align}
|\psi(t)>=c_{0}e^{-\frac{E_{0}}{\hbar}t}|\phi_{0}>+c_{1}e^{-\frac{E_{1}}{\hbar}t}|\phi_{1}>+\cdots
\end{align}
The ground state will decay slowest since it has the lowest energy. As a result, after every discrete imaginary time step, if we renormalize $|\psi(t)>$, it will have a larger and larger portion of $|\phi_{0}>$. When the energy coverages to a constant, we can consider $|\psi(t)>$ as $|\phi_{0}>$.

In this Hamiltonian, I noticed that there are differential operators that come from kinetic energy term $K$. In order to solve that, the fast Fourier transformation method instead of the time consuming finite difference method is applied. When $|\psi(x)>$ is transformed into Fourier space $|\psi(k)>$, $e^{-i\frac{K}{\hbar}t}$ can be multiplied directly since it's in the momentum space. After that I use the reverse transform to get $e^{-i\frac{K}{\hbar}t}|\psi(k)>$ back to real space and multiply it with $e^{-i\frac{V}{\hbar}t}$, which is the spatial dependent potential.

The algorithm for computing time evolution is
\begin{equation}
|\psi(x,t+\Delta t)>=e^{-i\frac{V}{\hbar}\Delta t}IFFT(e^{-i\frac{K}{\hbar}\Delta t}FFT(|\psi(x,t)>))
\end{equation}
Here $\Delta t$ is the discrete time step between two iterations.

\subsection{Spin-1/2 BEC without Interaction}
\begin{figure}
\includegraphics[width=0.5\textwidth]{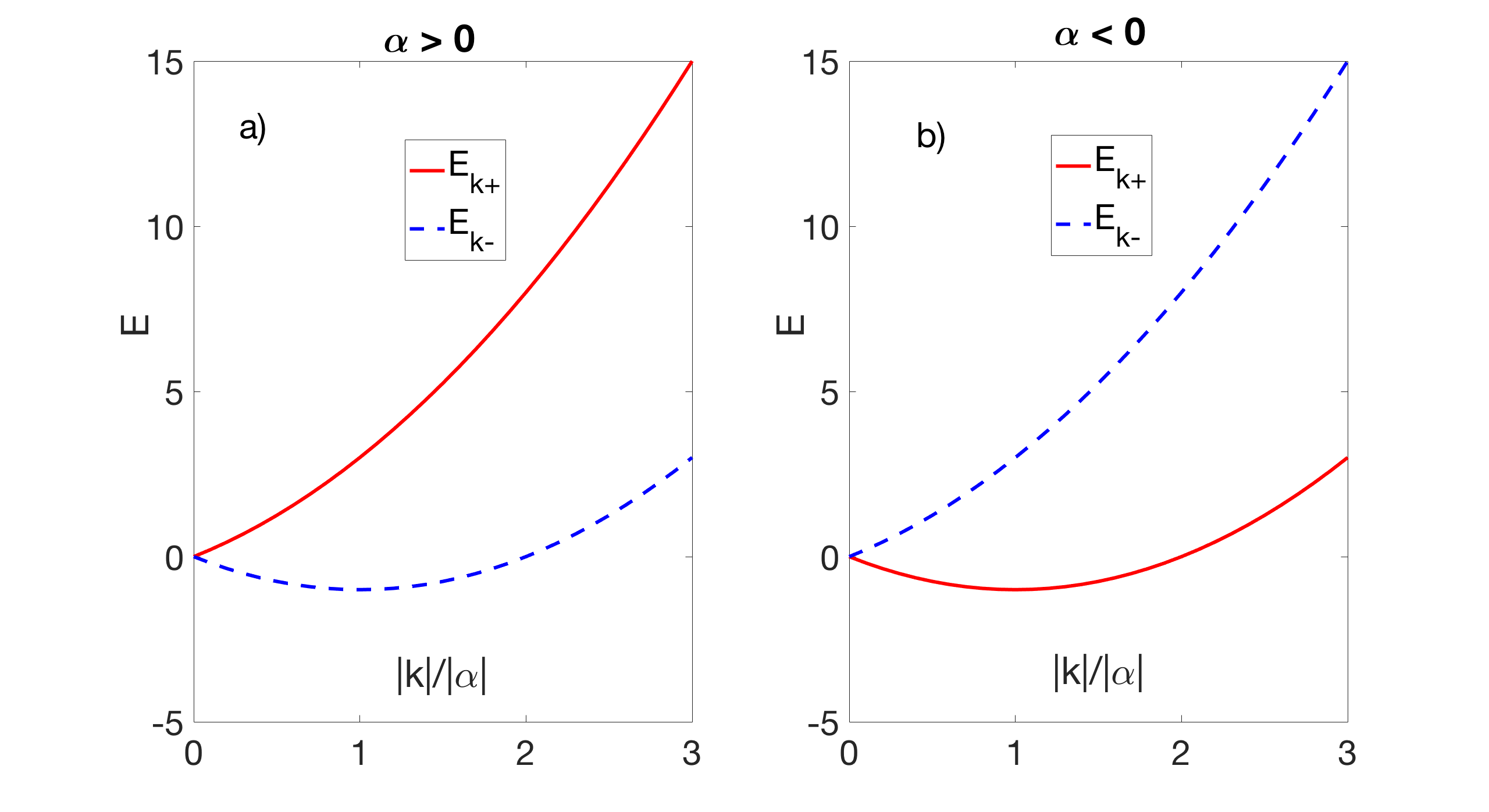}
\caption{(color online) The energy dispersion for helicity $\pm$ when a)  $\alpha > 0$ b) $\alpha <0$ }
\end{figure}

Hamiltonian of spin-1/2 BEC without interaction is identical to single particle Hamiltonian. 

Without external potential
\begin{equation}
H_{0}=\frac{1}{2m}(\vec{k}^{2}+2\alpha \vec{\sigma} \cdot \vec{k})
\end{equation}

Spin is no longer a good quantum number here. Instead, helicity is a good quantum number here. Helicity ``$\pm$'' means that the spin direction is either parallel or anti-parallel to the momentum. For these two helicity branches, their dispersion relation are given by
\begin{equation}
E_{k\pm}=\frac{1}{2m}(|k|^{2}\pm2\alpha |k|)
\end{equation}
where $|k|=\sqrt{k_{x}^{2}+k_{y}^{2}}$.

For the case $\alpha > 0$, the ``$-$'' helicity branch has lower energy. The single particle minimum is locating at $|k|=\alpha$. This minimum has degenerate ground states with different azimuthal angles. Without SO coupling, the Hamiltonian has a unique ground state at $k=0$. The weak interaction won't play a significant role here because the ground state is unique. However, in the presence of SO coupling the weak interaction plays a large part, since the ground states has a strong degeneracy.

Assuming the atoms are in a harmonic potential $V=\frac{1}{2}m\omega^{2}r^{2}$, since you need to trap the atom to observe it and one of the most common way is to use the dipole trap from a strong laser in experiment. In order to solve the Schr$\ddot{o}$dinger equation with the Hamiltonian (\ref{2nd quantized H}) for a many-body system, the mean-field theory is used here, which replaces the field operator with its expectation value $\phi_{m}=<\hat{\psi}_{m}>$. The energy of the system is in the form
 
 \begin{equation}
\begin{split}
E=\int d^{2} r \frac{1}{2m}(\sum_{m_{z}=\uparrow,\downarrow}\phi_{m_{z}}^{*}(k^{2}+\tilde{\omega}^{2}r^{2})\phi_{m_{z}}\\
+2\alpha \sum_{m_{z}, m_{z}'=\uparrow,\downarrow}\phi_{m_{z}}(\sigma_{x}k_{x}+\sigma_{y}k_{x})\phi_{m_{z}'})
\end{split}
\end{equation}
Here $\tilde{\omega}=m\omega$. Since what's important is the value of $\tilde{\omega}$ and $\alpha$, I neglect  the general factor $\frac{1}{2m}$ hereafter. 


For the spin-1/2 case, the Pauli matrices look like this:
\begin{equation}
\sigma_{x}=\begin{pmatrix}
0 & 1\\
1 & 0
\end{pmatrix},
\sigma_{y}=\begin{pmatrix}
0 & -i\\
i & 0
\end{pmatrix}, 
\sigma_{z}=\begin{pmatrix}
1 & 0 \\
0 & -1
\end{pmatrix}.
\end{equation}

Writing the energy in explicit form of spin components, gives: 
 \begin{equation}
\begin{split}
E=\int d^{2} r \{ \sum_{m_{z}=\uparrow,\downarrow}\phi_{m_{z}}^{*}(k^{2}+\tilde{\omega}^{2}r^{2})\phi_{m_{z}}\\
+2\alpha\{\phi_{\uparrow}^{*}(k_{x}-ik_{y})\phi_{\downarrow}+\phi_{\downarrow}^{*}(k_{x}+ik_{y})\phi_{\uparrow}\}.
\end{split}
\end{equation}

\begin{figure}
\includegraphics[width=0.5\textwidth]{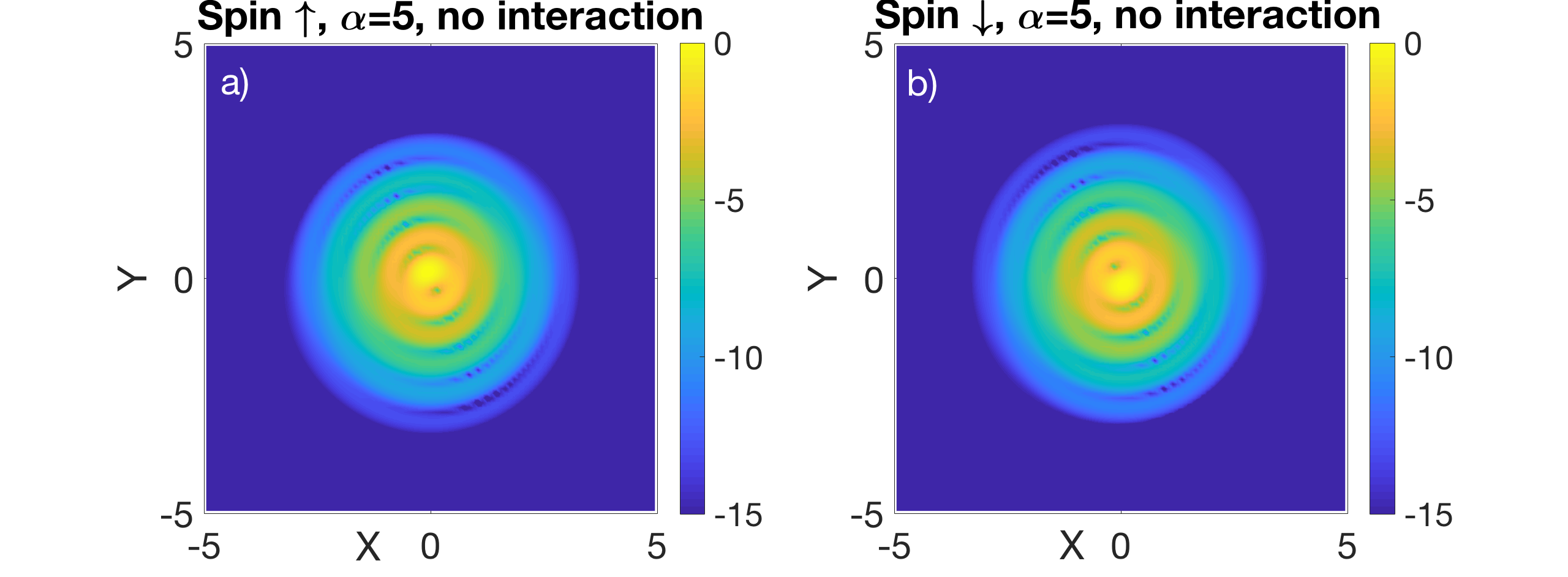}
\caption{(color online) The ground state density of both the spin up a) and down b) without scattering interaction, which has a strong degeneracy for different direction of $k$.}
\label{fig1}
\end{figure}

To solve it numerically, the main challenge comes from the two spin components ($\phi_{m_{z}'}^{*}A\phi_{m_{z}}, m_{z}'\neq m_{z}$) be coupled together. ($A$ is any operator). I can separated the coupled term $H_{so}$ from non coupled term $H_{0}$
\begin{equation}
\begin{split}
|\psi(t)>&=e^{-iH\Delta t}|\psi>\\
&=e^{-iH_{so}\Delta t}(e^{-iH_{0}\Delta t}I)|\psi>
\end{split}
\end{equation}
where $e^{-iH_{so}\Delta t}=e^{-i2\alpha(\sigma_{x}k_{x}+\sigma_{y}k_{y})\Delta t}$.

The $e^{-iH_{so}t}$ has an analytic formula that is solvable, because the exponential of a Pauli vector looks like:
\begin{equation}
e^{ia(\hat{n}\cdot \vec{\sigma})}= I \cos a+ i (\hat{n}\cdot \vec{\sigma})\sin a,
\end{equation}

\begin{equation}
e^{-i2\alpha(\sigma_{x}k_{x}+\sigma_{y}k_{y})\Delta t}=I \cos a+ i (\hat{n}\cdot \vec{\sigma})\sin a.
\end{equation}
Here, $a=-2\alpha\Delta t\sqrt{k_{x}^{2}+k_{y}^{2}}$ and $\hat{n}=(k_{x}/\sqrt{k_{x}^{2}+k_{y}^{2}}, k_{y}/\sqrt{k_{x}^{2}+k_{y}^{2}})$

\subsection{Spin-1/2 BEC with Interaction}

The interaction Hamiltonian term in a 2D spin-1/2 BEC is in the form of
\begin{equation}
H_{int}=\int d^{2}r \Psi^{\dagger} (\frac{c_{0}}{2}n^{2}+\frac{c_{2}}{2}S_{z}^{2}) \Psi
\end{equation}
$n$ is the operator of number of particles and $S_{z}$ is the operator of total spin $z$-component. This formula is derived from the contact interaction and the s-wave scattering. Now the time-dependent equation looks like

\begin{equation}
i\frac{\partial \psi}{\partial t}=(H_{0}+H_{SO}+H_{int})\psi
\end{equation}
which is called Gross-Pitaevskii equation \cite{Gross}.

 \begin{equation}
\begin{split}
E=\int d^{2} r \{ \sum_{m_{z}=\uparrow,\downarrow}\phi_{m_{z}}^{*}(-\nabla^{2}+\tilde{\omega}^{2}r^{2})\phi_{m_{z}}\\
+2\alpha\{\phi_{\uparrow}^{*}(k_{x}-ik_{y})\phi_{\downarrow}+\phi_{\downarrow}^{*}(k_{x}+ik_{y})\phi_{\uparrow}\}\\
+\frac{c_{0}}{2}(|\phi_{\uparrow}|^{2}+|\phi_{\downarrow}|^{2})^{2}+\frac{c_{2}}{2}(|\phi_{\uparrow}|^{2}-|\phi_{\downarrow}|^{2})^{2}
\end{split}
\end{equation}

Here below, all the numerical results are solved in atomic units (a.u.), which is $\hbar=1, m_{e}=1$.

The matrix form of the interaction Hamiltonian is:
\begin{widetext}
\begin{equation}
\begin{split}
&H_{int}=\\
&\begin{pmatrix}
\phi_{\uparrow}^{*}\\
\phi_{\downarrow}^{*}
\end{pmatrix}
\begin{pmatrix}
\frac{c_{0}+c_{2}}{2}|\phi_{\uparrow}|^{2}+\frac{c_{0}-c_{2}}{2}|\phi_{\downarrow}|^{2} & 0\\
0 & \frac{c_{0}+c_{2}}{2}|\phi_{\downarrow}|^{2}+\frac{c_{0}-c_{2}}{2}|\phi_{\uparrow}|^{2}
\end{pmatrix}
\begin{pmatrix}
\phi_{\uparrow}\\
\phi_{\downarrow}
\end{pmatrix}
\end{split}.
\end{equation}
\begin{equation}
\begin{split}
e^{-i H_{int}\Delta t}&=\\
&\begin{pmatrix}
e^{-i (\frac{c_{0}+c_{2}}{2}|\phi_{\uparrow}|^{2}+\frac{c_{0}-c_{2}}{2}|\phi_{\downarrow}|^{2})\Delta t} & 0\\
0 & e^{-i (\frac{c_{0}+c_{2}}{2}|\phi_{\downarrow}|^{2}+\frac{c_{0}-c_{2}}{2}|\phi_{\uparrow}|^{2})\Delta t}
\end{pmatrix}
\end{split}
\end{equation}

\end{widetext}

To simplify the expression, I use $\gamma=c_{2}/c_{0}$ as the ratio between the different scattering lengths instead of using $c_{0}$, $c_{2}$ as separate coefficients.
 
Combining these terms, I formulate the detailed equation that will be used to calculate the ground state.

\begin{equation}
\begin{split}
&\psi(t+\Delta t)=e^{-i (H_{0}+H_{SO}+H_{int})\Delta t}\psi(t)\\
&=e^{-i H_{SO}\Delta t}e^{-i (H_{0}+H_{int})\Delta t}\psi(t)\\
&=(I \cos a+ i (\hat{n}\cdot \vec{\sigma})\sin a)e^{-i (H_{0}+H_{int})\Delta t}\psi(t)
\end{split}
\end{equation}

With:

\begin{equation}
\begin{split}
&e^{-i H_{0}\Delta t}=\\
&\begin{pmatrix}
e^{-i(k^{2}+\tilde{\omega}^{2}r^{2})\Delta t} & 0\\
0 & e^{-i(k^{2}+\tilde{\omega}^{2}r^{2})\Delta t}
\end{pmatrix}
\end{split}
\end{equation}
\begin{equation}
\begin{split}
&I \cos a+ i (\hat{n}\cdot \vec{\sigma})\sin a=\\
&\begin{pmatrix}
\cos a & i(n_{x}-in_{y})\sin a\\
i(n_{x}+in_{y})\sin a & \cos a
\end{pmatrix}.
\end{split}
\end{equation}

\begin{figure}
\includegraphics[width=0.5\textwidth]{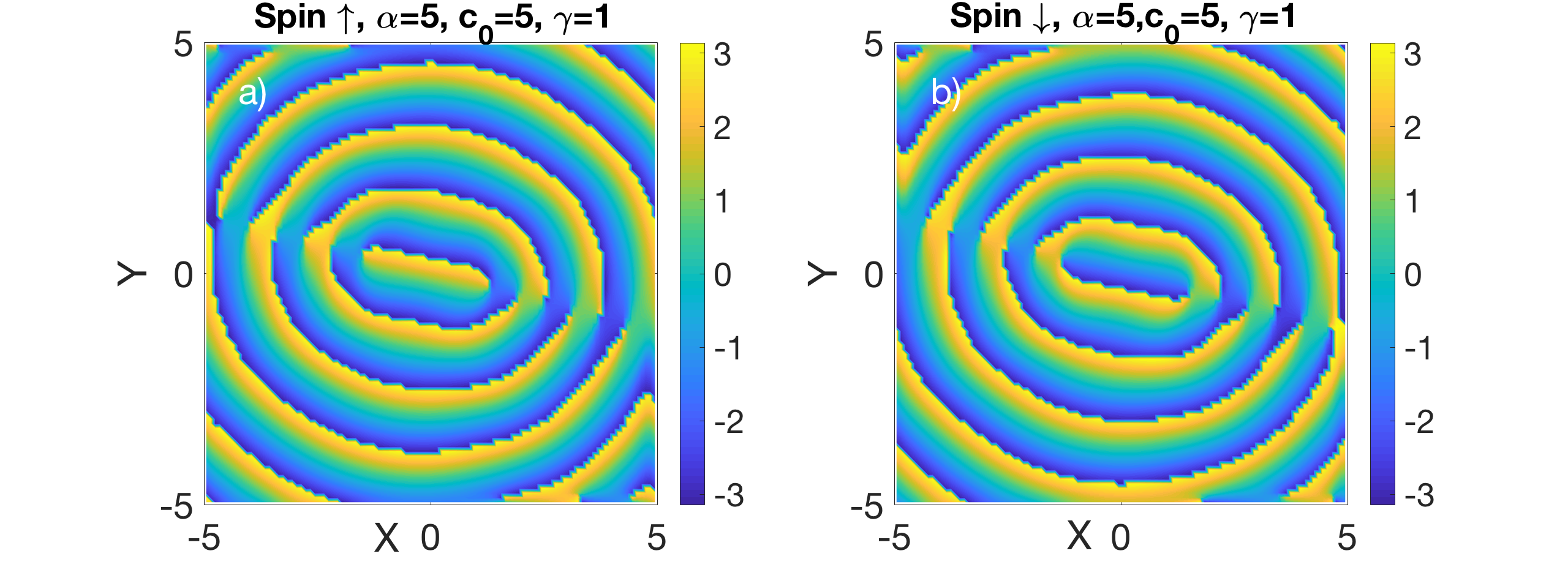}
\caption{(color online) The phase of both the spin $\uparrow$ a) and $\downarrow$ b)  ground state with scattering interaction $c_{0}=5,\gamma=1$. Since the phase increase from $-\pi$ (dark regime) to $\pi$ (bright regime) periodically, we call it ''plane wave phase'' (PW) }
\label{PW_phase}
\end{figure}

\begin{figure}
\includegraphics[width=0.5\textwidth]{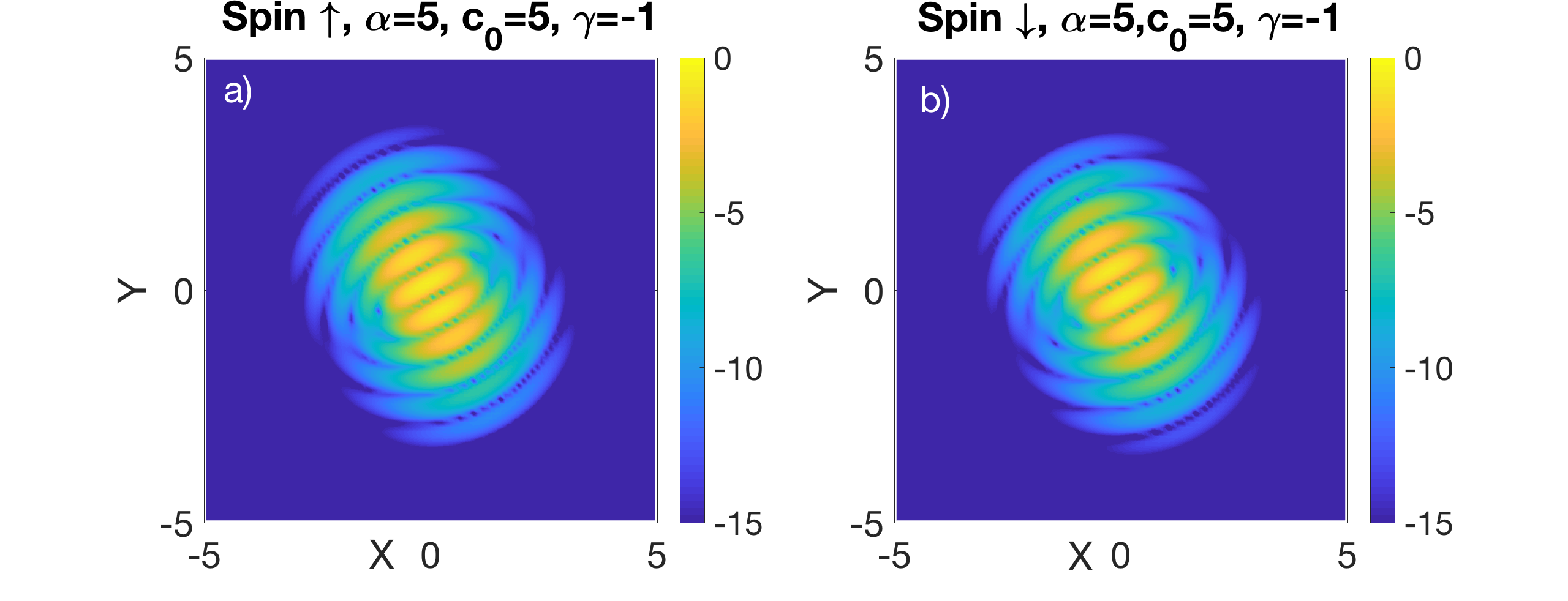}
\caption{(color online) The ground state density of both the spin $\uparrow$ a) and $\downarrow$ b) with scattering interaction $c_{0}=5,\gamma=-1$, which is the SW phase. From the figures we can see both components oscillate periodically in the SW regime.}
\label{SW}
\end{figure}

As a result, we can also solve the equation to get the ground state for the BEC with interaction. When $\gamma>0$, the ground state is called a plane wave (PW) phase. This is because the phase of condensate wave function looks like a plane wave in Fig. \ref{PW_phase}. When $\gamma<0$, the ground state is called standing wave (SW) phase. It's pretty straightforward to understand it because the density looks pretty much like a standing wave (Fig. \ref{SW}) \cite{SOtheory}.

Although in numerical simulation, the harmonic trap is included to avoid an artifact from a sharp boundary and also simulate the practical situation of a cold atom experiment, the results can be understood without external potential. With SO coupling, $E_{\pm k}=|k|^{2}/2\pm\alpha |k|$ from the single particle Hamiltonian discussion above. Here the ``$\pm$'' denotes different helicity (spin orthogonal to wave factor). When $|\alpha|>0$, the single particle ground state is in the negative helicity branch, with $|k|=\alpha$ by minimizing the energy. The wave function is
\begin{equation}
\phi_{k}=\frac{1}{\sqrt{2}}e^{{ikr}}\begin{pmatrix}
1\\ e^{i\varphi_{k}}
\end{pmatrix}
\end{equation}
where $\varphi_{k}=arg(k_{x}+ik_{y})+\pi$ which is anti-parallel to momentum, and $\varphi_{-k}=\varphi_{k}+\pi$.

\begin{figure}
\includegraphics[width=0.5\textwidth]{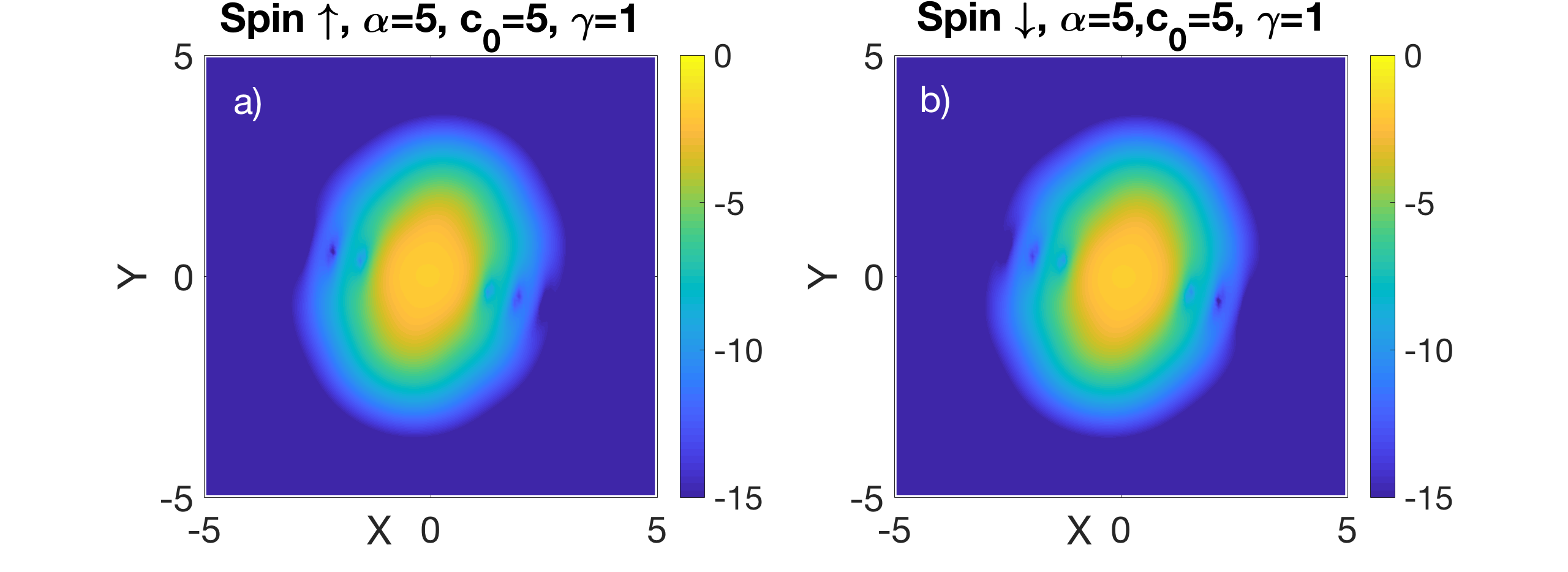}
\caption{(color online) The ground state density of both the spin $\uparrow$ a) and $\downarrow$ b) with scattering interaction $c_{0}=5,\gamma=1$, which is the PW phase. From the figures we can see the vortices work as the domain wall.}
\label{PW}
\end{figure}

Consider the wave function as a composition of two opposite wave vector states as 
\begin{equation}
\varphi=\frac{\alpha_{1}}{\sqrt{2}}e^{ikr}\begin{pmatrix}
1\\ e^{i\varphi_{k}}\end{pmatrix}+\frac{\alpha_{2}}{\sqrt{2}}e^{-ikr}\begin{pmatrix}
1\\ e^{i\varphi_{-k}}\end{pmatrix}.
\end{equation}

Using this wave function to minimize the interaction energy. When $ \gamma>0 $, we get $\alpha_{1}=1, \alpha_{2}=0 \text{ or } \alpha_{1}=0, \alpha_{2}=1$.
\begin{equation}
\varphi=\frac{1}{\sqrt{2}}e^{ikr}\begin{pmatrix}
1\\ e^{i\varphi_{k}}\end{pmatrix} \text{ or }\frac{1}{\sqrt{2}}e^{-ikr}\begin{pmatrix}
1\\ e^{i\varphi_{-k}}\end{pmatrix}
\end{equation}
The wave function is a single plane wave, corresponding to the PW phase. This state breaks the time-reversal, the rotational symmetry and the U(1) symmetry of the superfluid phase.

When $\gamma<0$, we get $\alpha_{1}=\alpha_{2}=1/\sqrt{2}$.
\begin{equation}
\varphi=\frac{1}{2}e^{ikr}\begin{pmatrix}
1\\ e^{i\varphi_{k}}\end{pmatrix}+\frac{1}{2}e^{-ikr}\begin{pmatrix}
1\\ e^{i\varphi_{-k}}\end{pmatrix}
\end{equation}

Thus
\begin{equation}
\phi_{\uparrow}\propto \cos(kr), \phi_{\downarrow}\propto i\sin(kr)
\end{equation}
Which corresponds to the SW phase (also named as a ``stripe superfluid''). This state breaks rotational symmetry while keeping the U(1) symmetry of the superfluid phase, reflection symmetry and translation symmetry along stripe direction.

\begin{figure}
\includegraphics[width=0.5\textwidth]{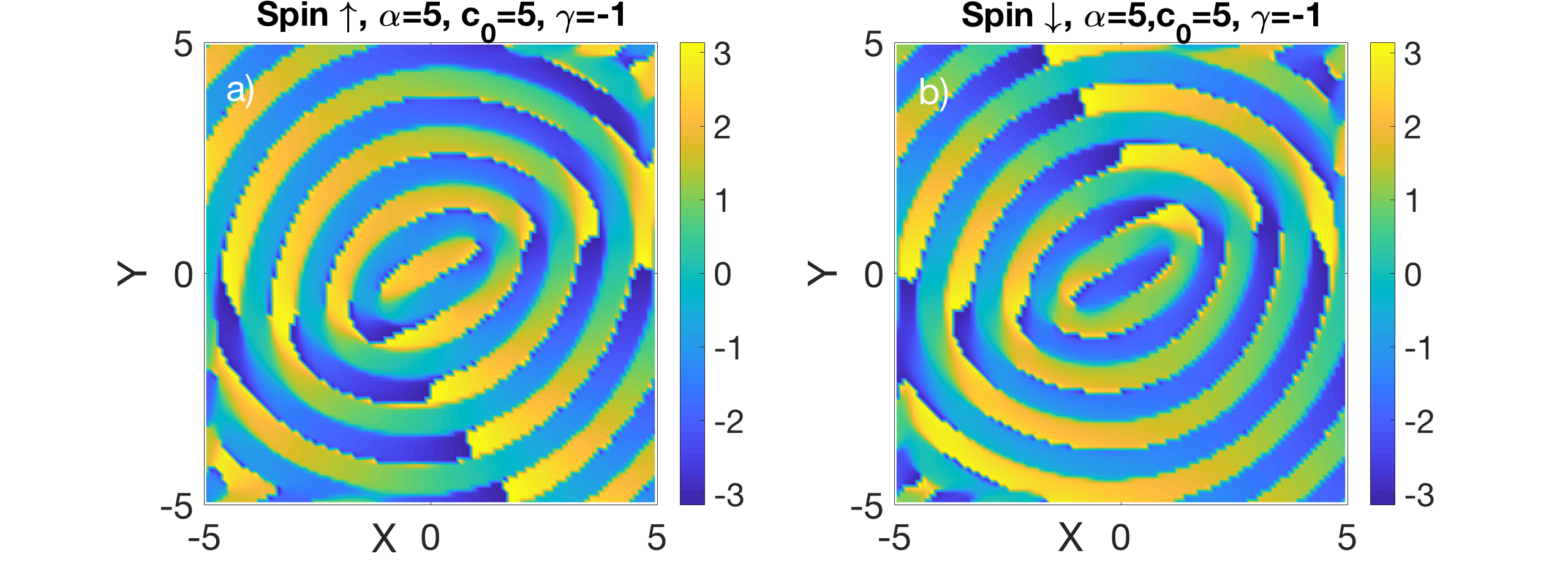}
\caption{(color online) The phase of both the spin $\uparrow$ a) and $\downarrow$ b)  ground state with scattering interaction $c_{0}=5,\gamma=-1$. }
\label{SW_phase}
\end{figure}

In general, we should substitute the single $k$ state with a superposition of all wave vector states with different azimuthal angles: 

\begin{equation}
\int d\varphi_{k} \frac{\alpha_{k}}{\sqrt{2}}e^{ikr}\begin{pmatrix}
1\\ e^{i\varphi_{k}}\end{pmatrix}
\end{equation}

If we minimize the energy with respect to $\alpha_{k}$, we can find the most favorable solution is alway single k or for a pair of $\{k,-k\}$. The reason of the symmetry breaking is because the interaction term has preference in specific spin vectors. This shows how a nearly can isotropic SO coupling enhance the effects of inter-particle interactions.

\subsection{Spin-1 BEC without Interaction}
We can also derive the Hamiltonian for the spin-1 case. For this case, the Pauli matrices look like this:
\begin{equation}
\begin{split}
&\sigma_{x}=\frac{1}{\sqrt{2}}\begin{pmatrix}
0 & 1 & 0\\
1 & 0 & 1\\
0 & 1 & 0
\end{pmatrix},
\sigma_{y}=\frac{i}{\sqrt{2}}\begin{pmatrix}
0 & -1 & 0\\
1 & 0 & -1\\
0 & 1 & 0
\end{pmatrix}, \\
&\sigma_{z}=\begin{pmatrix}
1 & 0 & 0\\
0 & 0 & 0\\
0 & 0 & -1
\end{pmatrix}
\end{split}
\end{equation}

The energy now changes into the form of 
 \begin{equation}
\begin{split}
E=\int d^{2} r \{ \sum_{m_{z}=1,0,-1}\phi_{m_{z}}^{*}(-\frac{\hbar^{2}}{2m}\nabla^{2}+\frac{1}{2}m\omega^{2}r^{2})\phi_{m_{z}}\\
+\alpha/\sqrt{2}\{\phi_{1}^{*}(k_{y}+ik_{x})\phi_{0}+\phi_{0}^{*}(k_{y}-ik_{x})\phi_{1}\\
+\phi_{0}^{*}(k_{y}+ik_{x})\phi_{-1}+\phi_{-1}^{*}(k_{y}+ik_{x})\phi_{0}\}
\end{split}
\end{equation}

 \begin{equation}
\begin{split}
E=\int d^{2} r \{ \sum_{m_{z}=1,0,-1}\phi_{m_{z}}^{*}(-\frac{\hbar^{2}}{2m}\nabla^{2}+\frac{1}{2}m\omega^{2}r^{2})\phi_{m_{z}}\\
+\alpha/\sqrt{2}\{\phi_{1}^{*}(i\partial_{y}-\partial_{x})\phi_{0}+\phi_{0}^{*}(i\partial_{y}+\partial_{x})\phi_{1}\\
+\phi_{0}^{*}(i\partial_{y}-\partial_{x})\phi_{-1}+\phi_{-1}^{*}(i\partial_{y}+\partial_{x})\phi_{0}\}
\end{split}
\end{equation}

\begin{figure}
\includegraphics[width=0.5\textwidth]{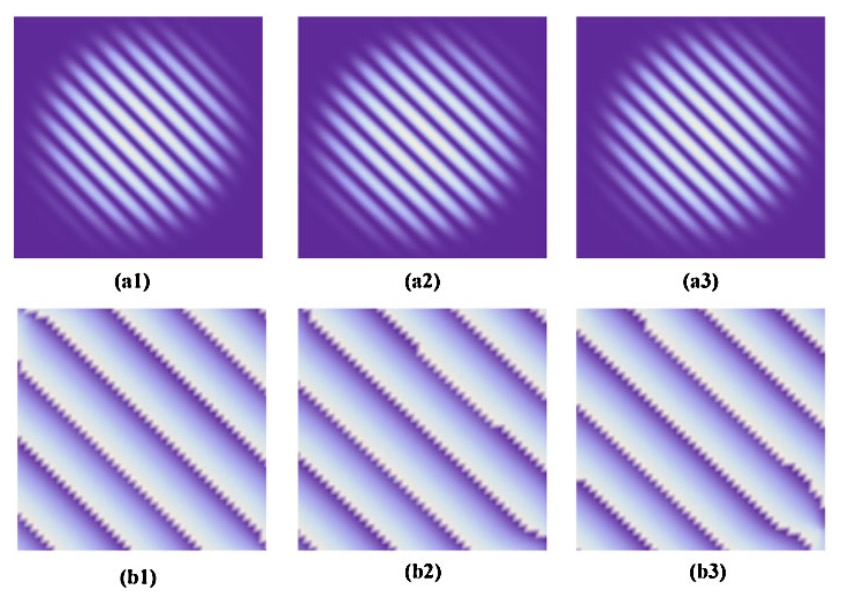}
\caption{(color online) Numerical results for the spin-1 cases. a1-a3 are density of -1, 0, 1 components in the SW phase with $\gamma=0.2$. b1-b3 are phase of -1, 0, 1 components in the PW phase with $\gamma=-0.2$. \cite{SOtheory}}
\label{spin-1}
\end{figure}

This is  also solvable. This case is a little bit complicated to program, so here I just present the results in Fig. \ref{spin-1} from \cite{SOtheory}.

\section{Relevant Experiment}
\begin{figure}
\includegraphics[width=0.2\textwidth]{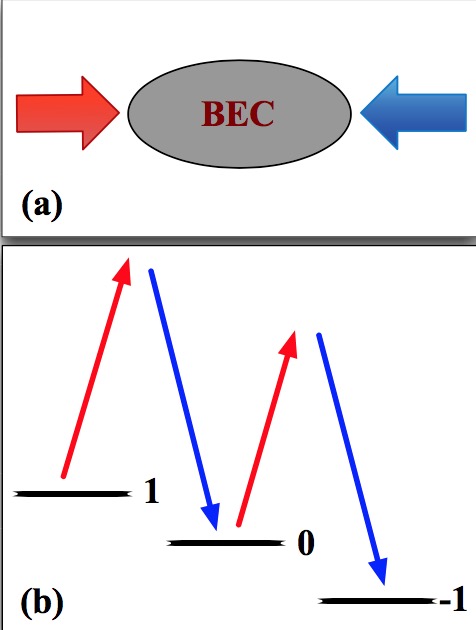}
\caption{(color online) a) A schematic of NIST experiment, in which two counter propagating Raman beams are applied. b) A schematic of how $F=1$ levels are coupled by Raman beams. \cite{review}}
\label{exp_setup}
\end{figure}

Considering an experiment with a $^{87}\text{Rb}$ Bose-Einstein condensate (BEC), where a pair of Raman lasers create a momentum-sensitive coupling between two internal atomic states \cite{spielman}. This SO coupling is equivalent to that of an electronic system with equal distribution of Rashba and Dresselhaus couplings, and with a uniform magnetic field of $B$ in the $y-z$ plane. The derivation of the Hamiltonian is shown below.

The Raman resonance is a phenomenon which requires two-photon processes. As Fig. \ref{exp_setup} a) shows, I have two counter-propogating beams shine on BEC.  The blue beam is $\sqrt{\Omega/2} e^{i(\omega_{1}t+k_{1}x)}$ while the red beam is $\sqrt{\Omega/2} e^{i(\omega_{2}t-k_{2}x)}$. The frequency difference $\omega_{1}-\omega_{2}=\delta \omega$. Here $\delta \omega$ is the energy gap between two spin components. For this reason, the coupling term becomes:
\begin{equation}
\begin{split}
H_{12}=\sqrt{\Omega/2} e^{i(\omega_{1}t+k_{1}x)}\sqrt{\Omega/2} e^{i(\omega_{2}t-k_{2}x)}\\
=\Omega/2 e^{i(\delta \omega t + (k_{1}+k_{2})x)}
\end{split}
\end{equation}
At the situation where $\delta \omega \gg k_{1}, k_{2}$ and $k_{1}\approx k_{2}=k_{0}$, the coupling term becomes $H_{12}=\Omega e^{i2k_{0}x}$.
The matrix form in the bases of $|1,-1>, |1,0>$
\begin{equation}
H=\begin{pmatrix}
\frac{k_{x}^{2}}{2m}+\frac{h}{2} & \frac{\Omega}{2}e^{i2k_{0}x}\\
\frac{\Omega}{2}e^{-i2k_{0}x} & \frac{k_{x}^{2}}{2m}-\frac{h}{2}
\end{pmatrix}
\end{equation}

By applying a unitary transformation with 
\begin{equation}
U=\begin{pmatrix}
e^{-ik_{0}x} & 0\\
0 & e^{ik_{0}x}
\end{pmatrix}
\end{equation}

One reaches an effective Hamiltonian that describes SO coupling
\begin{equation}
\begin{split}
&H_{SO}=UH_{SO}U^{\dagger}\\
&=\begin{pmatrix}
\frac{(k_{x}+k_{0})^{2}}{2m}+\frac{h}{2} & \frac{\Omega}{2}\\
\frac{\Omega}{2} & \frac{(k_{x}-k_{0})^{2}}{2m}-\frac{h}{2}
\end{pmatrix}\\
&=\frac{1}{2m}(k_{x}+k_{0}\sigma_{z})^{2}+\frac{\Omega}{2}\sigma_{x}+\frac{h}{2}\sigma_{z}
\end{split}
\end{equation}

Using the pseudo-spin rotation $\sigma_{x}'=-\sigma_{z}, \sigma_{z}'=\sigma_{x}$, and combining all constants in a single coefficient, we have
\begin{equation}
H=\frac{1}{2m}(k^{2}+\Omega\sigma_{z}+\delta\sigma_{x}+\alpha k_{x}\sigma_{x})
\end{equation}
In this Hamiltonian, we get the last term $k_{x}\sigma_{x}$ equivalent to an equal distribution of Rashba and Dresselhaus couplings. Thus, this Raman laser coupling system is equivalent to a 2D SO coupling: 

\begin{equation}
H=\frac{\hbar^{2}\vec{k}^{2}}{2m}-(\vec{B}+\vec{B_{SO}}(k))\cdot \vec{\mu}.
\end{equation}

Since in realty, there is no spin $1/2$ bosons, we use $F=1$ ground electronic manifold and label them with pseudo-spin-up and pseudo-spin-down: $|\uparrow>=|F=1,m_{F}=0>$ and $|\downarrow>=|F=1,m_{F}=-1>$. The initial BEC state is an equal population of $|\uparrow>$ and $|\downarrow>$.

\begin{figure}
\includegraphics[width=0.5\textwidth]{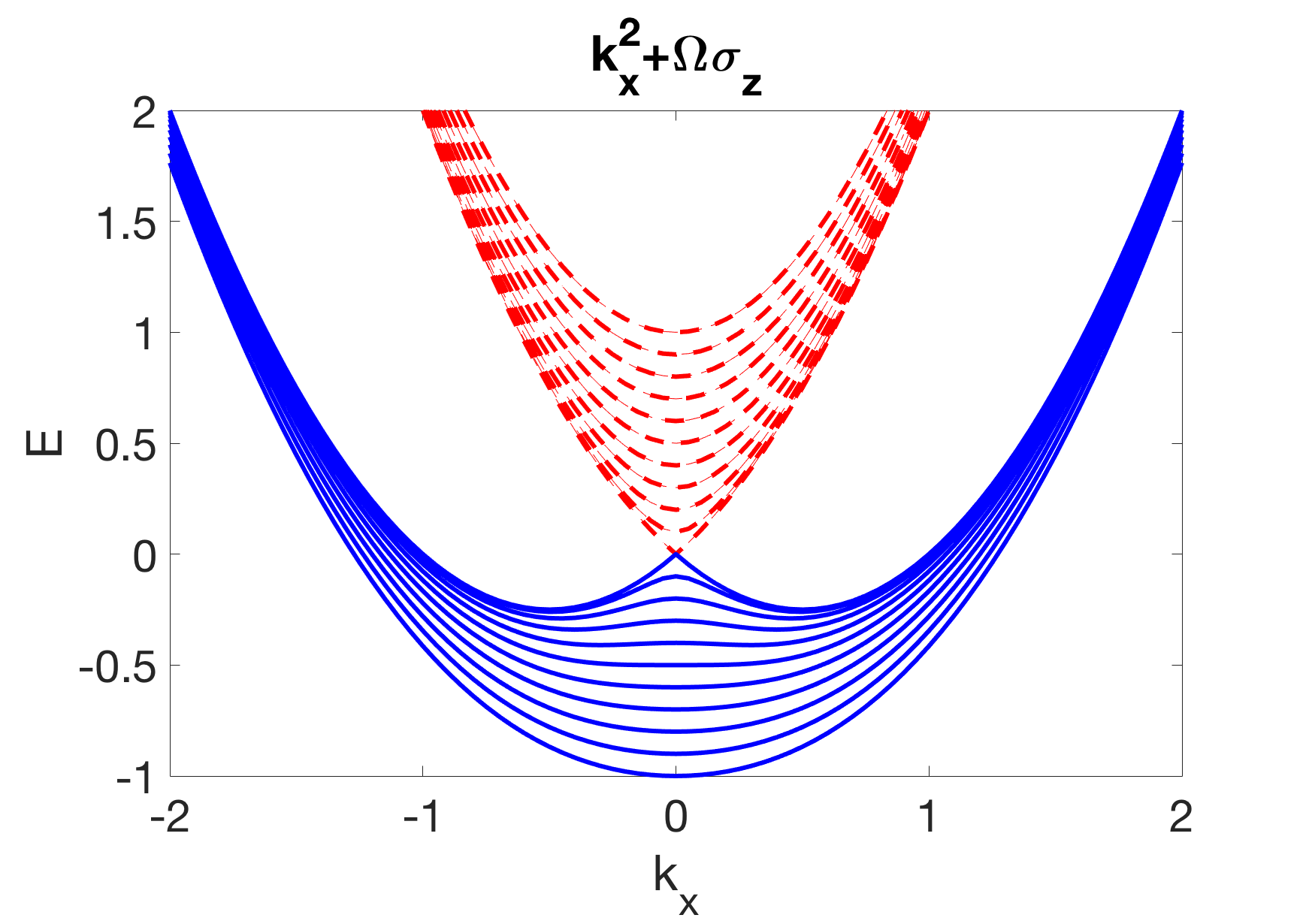}
\caption{(color online) The dispersion relation for the Hamiltonian in the experiment with $\delta=0$ and $\Omega=0:0.1:1$.}
\label{disperse_relation_exp}
\end{figure}

Let's do the same analysis, as the theory I mentioned previously.  Without the interaction, we can get the dispersion relation

\begin{equation}
E=\frac{1}{2m}(k_{x}^{2}\pm\sqrt{\Omega^{2}+\alpha^{2} k_{x}^{2}+2\alpha\delta k_{x}+\delta^{2}})
\end{equation} 
where the eigenstates are
\begin{equation}
u=\begin{pmatrix}
\frac{\Omega-\sigma_{1}}{\delta + \alpha k_{x}}\\
1
\end{pmatrix}, d=\begin{pmatrix}
\frac{\Omega+\sigma_{1}}{\delta + \alpha k_{x}}\\
1
\end{pmatrix}
\end{equation}
Here $\sigma_{1}=\sqrt{\Omega^{2}+\alpha^{2}k_{x}^{2}+2\alpha\delta k_{x}+\delta^{2}}$.

When $\delta=0$ and I increase $\Omega$, the dispersion relation changes from two minima at $k_{x}\neq 0$ to a single minimum $k_{x}=0$ as shown in Fig. \ref{disperse_relation_exp}.

Using the same numerical method as mentioned before, the ground state can be calculated, here the $|\uparrow>_{x}, |\downarrow>_{x}$ are spin up and down eigenstates in the $x$ direction. It's convenient to use this basis because of the two $\sigma_{x}$ terms in Hamiltonian.

\begin{figure}
\includegraphics[width=0.5\textwidth]{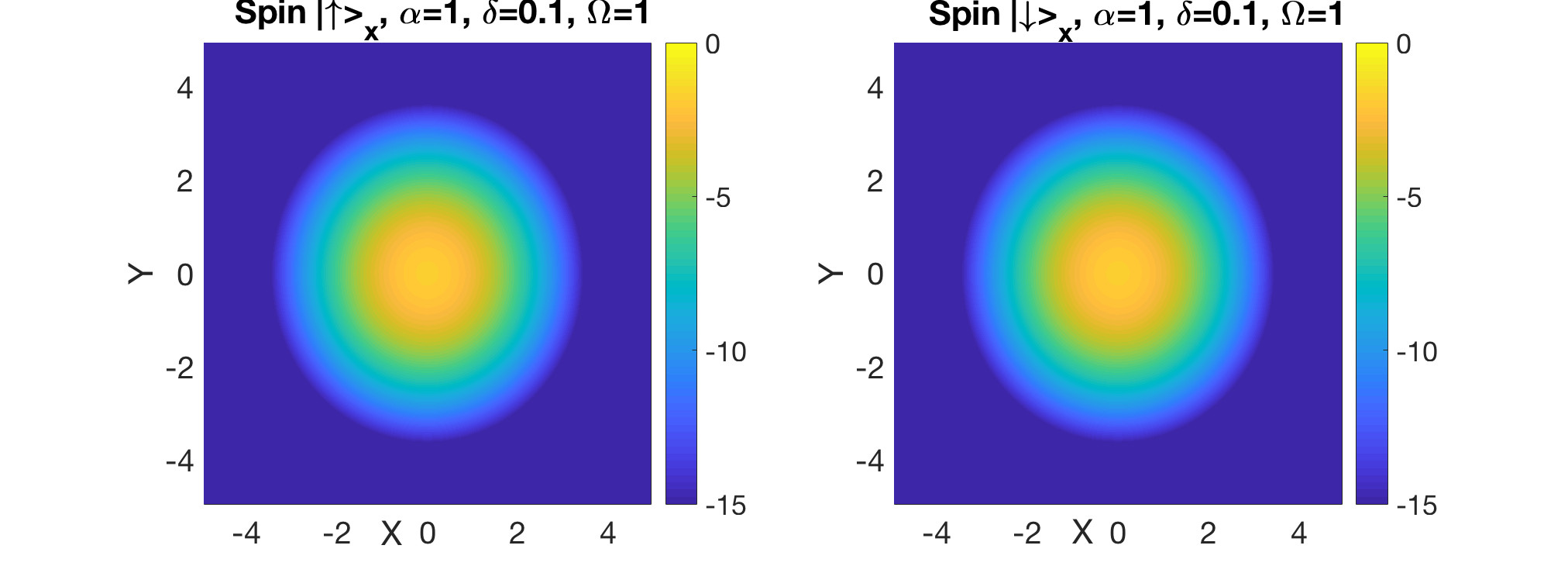}
\caption{(color online) The ground state density when $\Omega$ is large, so there is only one minima and it has an equal distribution in $|\uparrow>_{x}, |\downarrow>_{x}$.}
\label{}
\end{figure}

\begin{figure}
\includegraphics[width=0.5\textwidth]{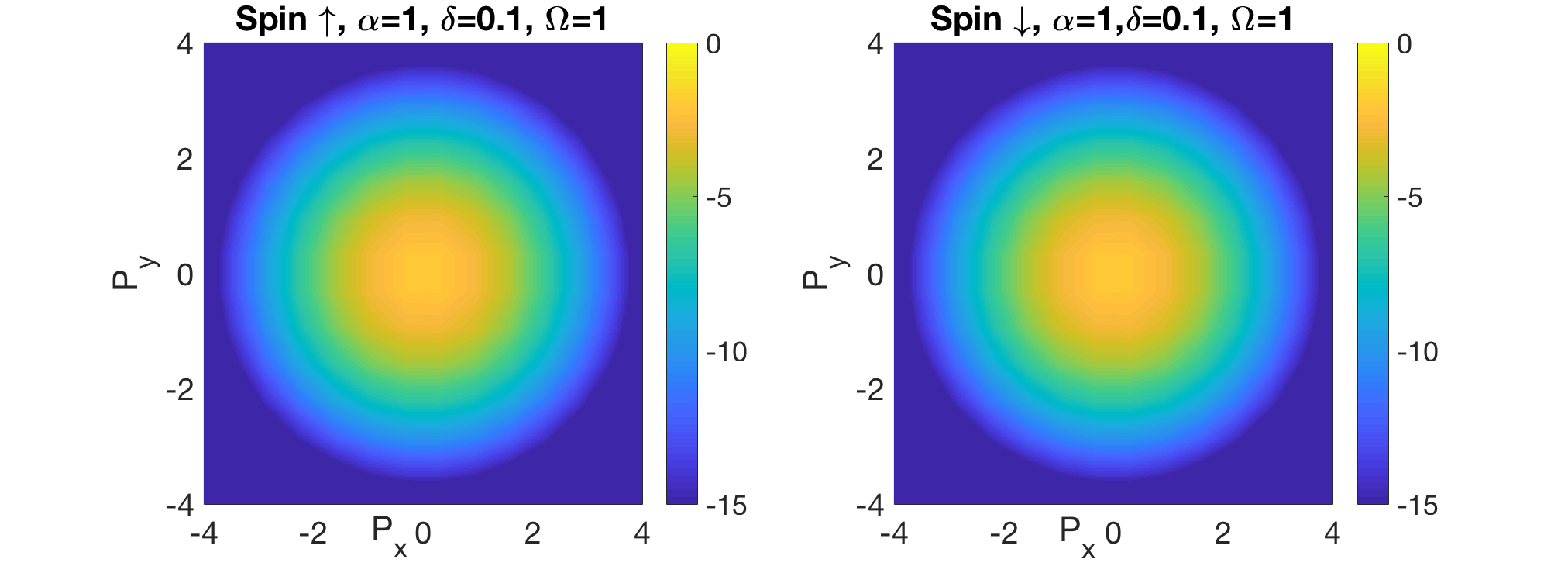}
\caption{(color online) The ground state momentum density when $\Omega$ is large, since there is only one minima at $k_{x}=0$. It's easy to see the momentum density is centered at $k_{x}=0$.}
\label{}
\end{figure}
 
\begin{figure}
\includegraphics[width=0.38\textwidth]{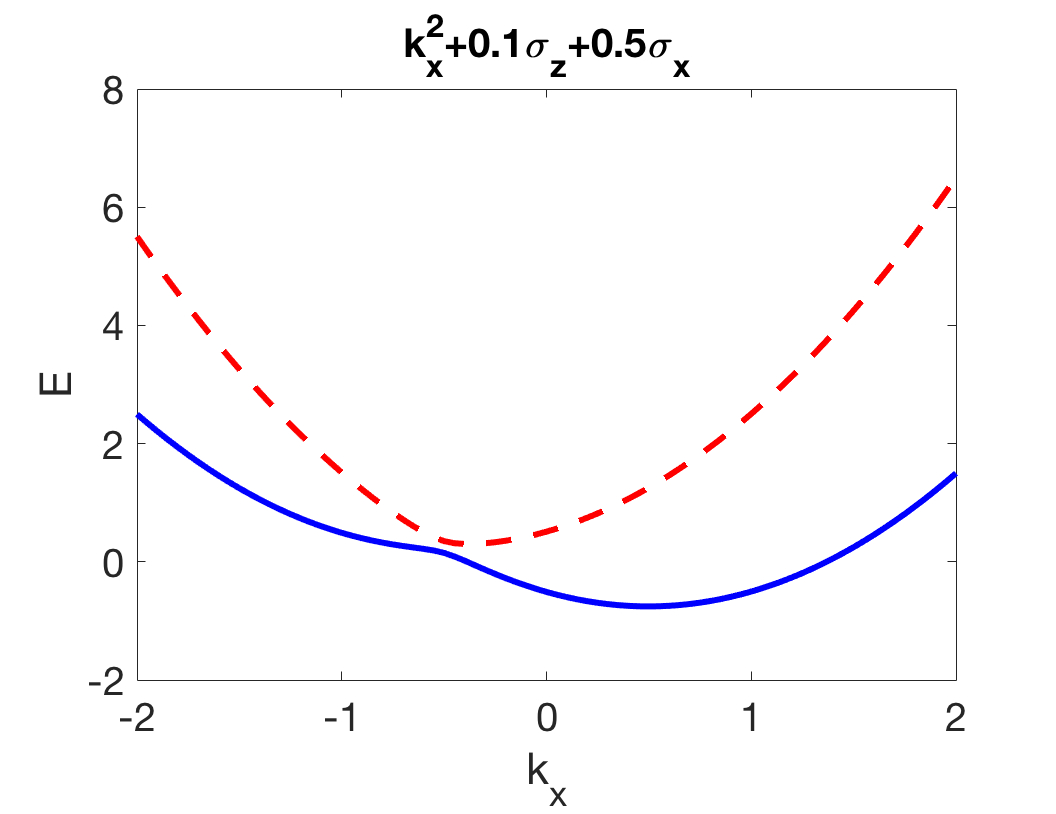}
\caption{(color online) The dispersion relation for the Hamiltonian in the experiment with small positive $\delta$, which tilt the ground state has two local minimum $|\downarrow_{x},k_{x}>0>, |\uparrow_{x},k_{x}<0>$.}
\label{spin_down_phase_dispersion}
\end{figure}

\begin{figure}
\includegraphics[width=0.5\textwidth]{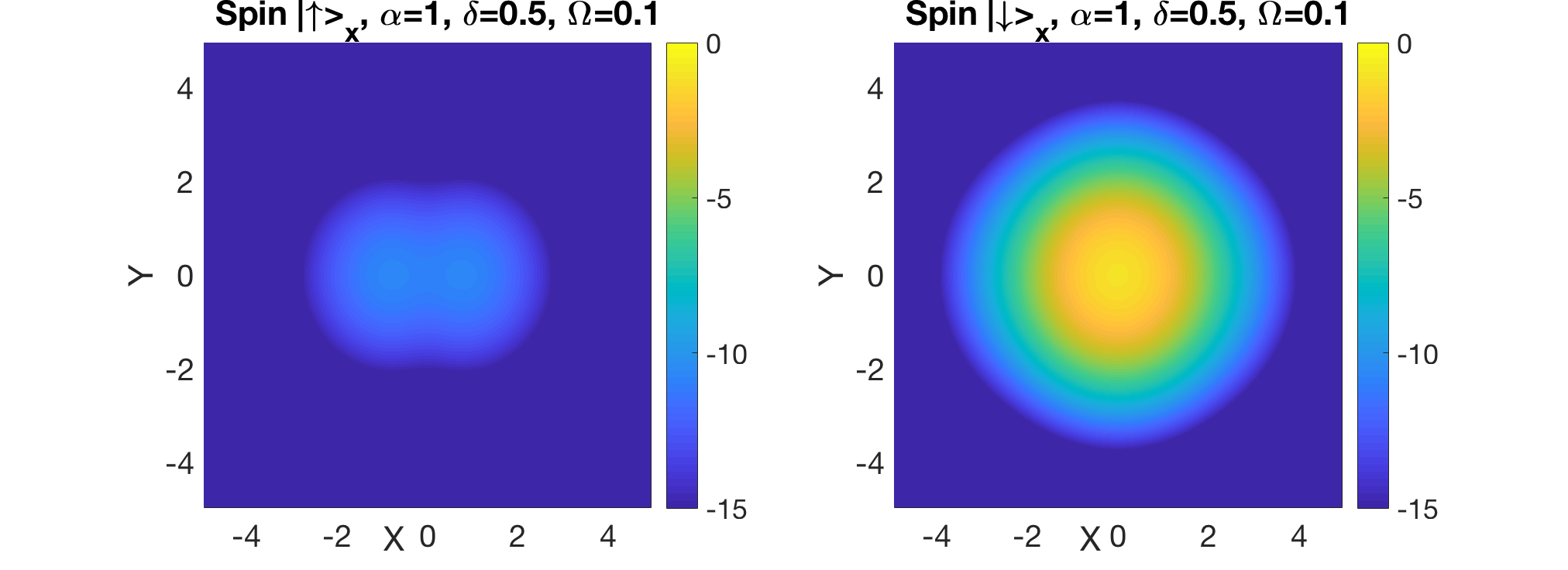}
\caption{(color online) The ground state density for small positive $\delta$. Large fraction of ground state is in $|\downarrow>_{x}$ as shown in the Fig. \ref{spin_down_phase_dispersion}.}
\label{spin_down_phase}
\end{figure}

\begin{figure}
\includegraphics[width=0.5\textwidth]{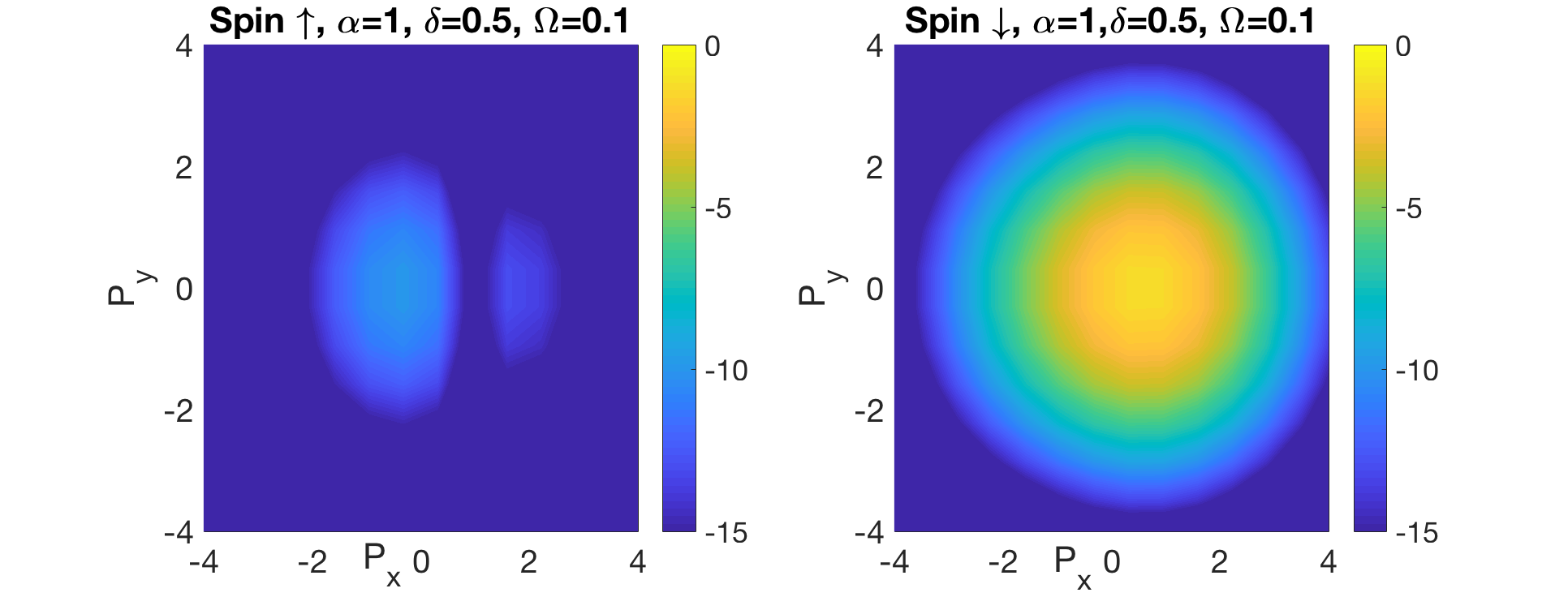}
\caption{(color online) The ground state momentum density when $\delta$ is positive, since there are two local minimum at $k_{x}\neq 0$. Comparing to Fig. \ref{spin_down_phase}, the pairs $|\downarrow_{x},k_{x}>0>, |\uparrow_{x},k_{x}<0>$ are indicated clearly.}
\label{spin_down_phase_p}
\end{figure}

When $\Omega=0$ and I change $\delta > 0$, the dispersion relation no longer has two global minima instead one of them changes into two local minima as shown in Fig. \ref{spin_down_phase_dispersion}. Thus, the ground state will prefer the $|\downarrow>_{x}$ components and change the average $k_{x}>0$. If $\Omega$ is not very strong, there will still be some fraction in $|\uparrow_{x},k_{x}<0>$. This result is shown in Fig. \ref{spin_down_phase} and Fig. \ref{spin_down_phase_p}.

\begin{figure}
\includegraphics[width=0.4\textwidth]{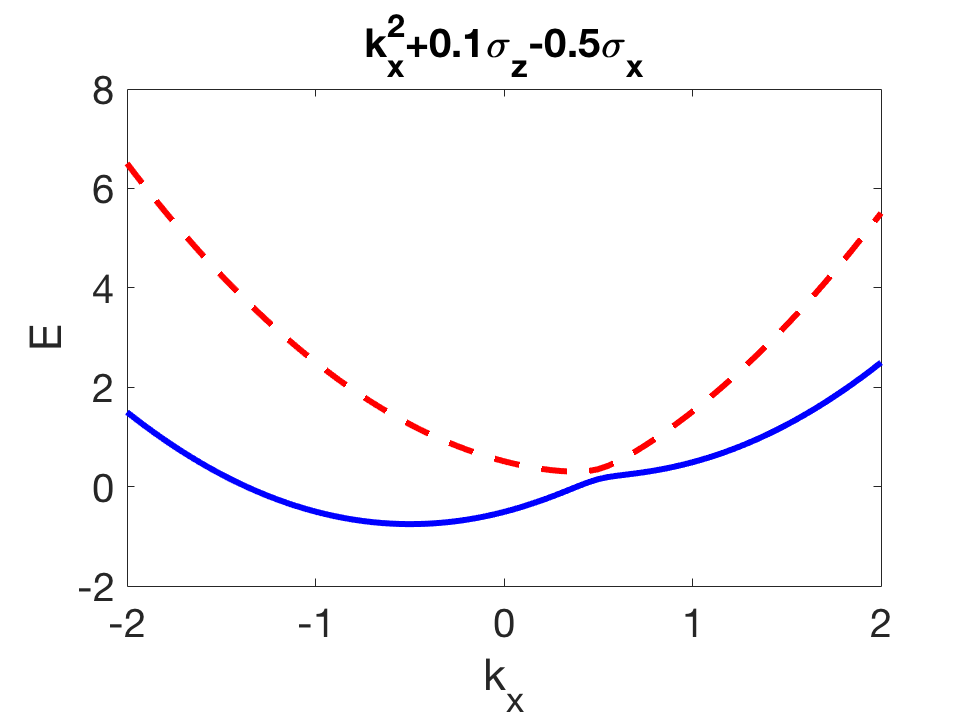}
\caption{(color online) The dispersion relation for the Hamiltonian in the experiment with small negative $\delta$, which tilt the ground state has two local minimum $|\downarrow_{x},k_{x}>0>, |\uparrow_{x},k_{x}<0>$.}
\label{spin_up_phase_dispersion}
\end{figure}

\begin{figure}
\includegraphics[width=0.5\textwidth]{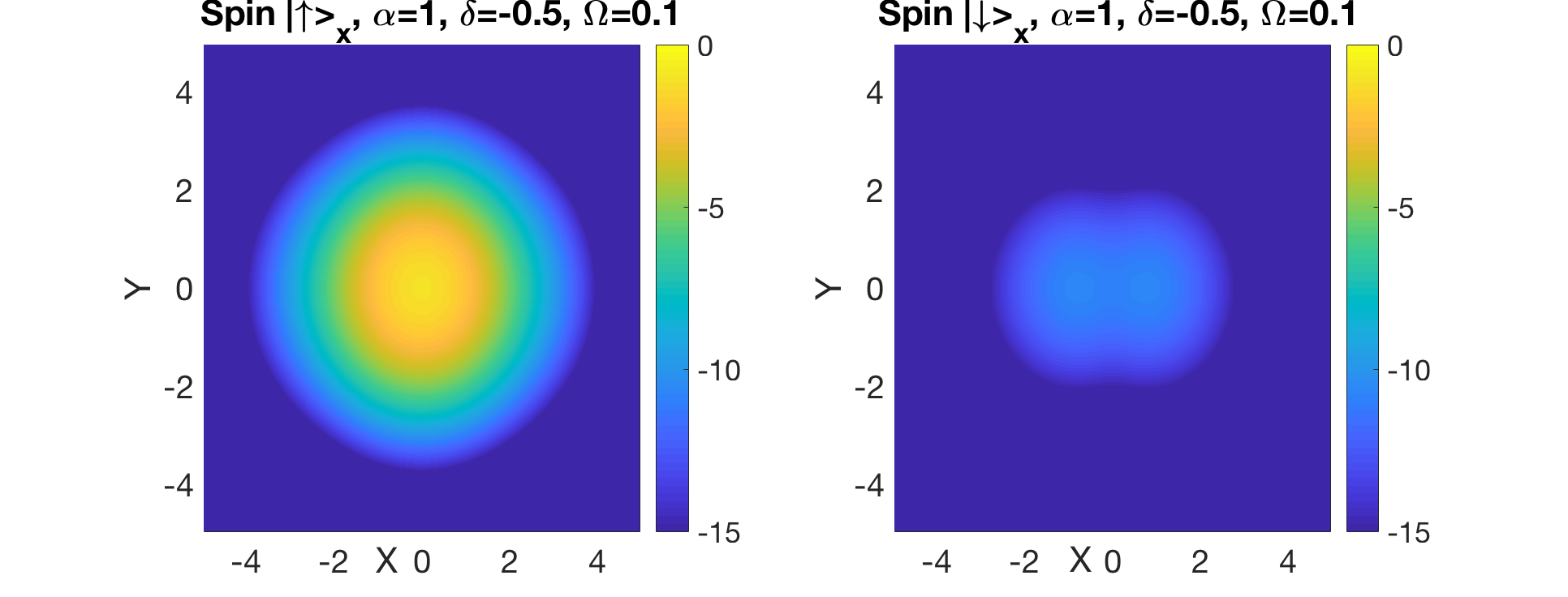}
\caption{(color online) The ground state density for small negative $\delta$. Large fraction of ground state is in $|\uparrow>_{x}$ as shown in the Fig. \ref{spin_up_phase_dispersion}.}
\label{spin_up_phase}
\end{figure}

\begin{figure}
\includegraphics[width=0.5\textwidth]{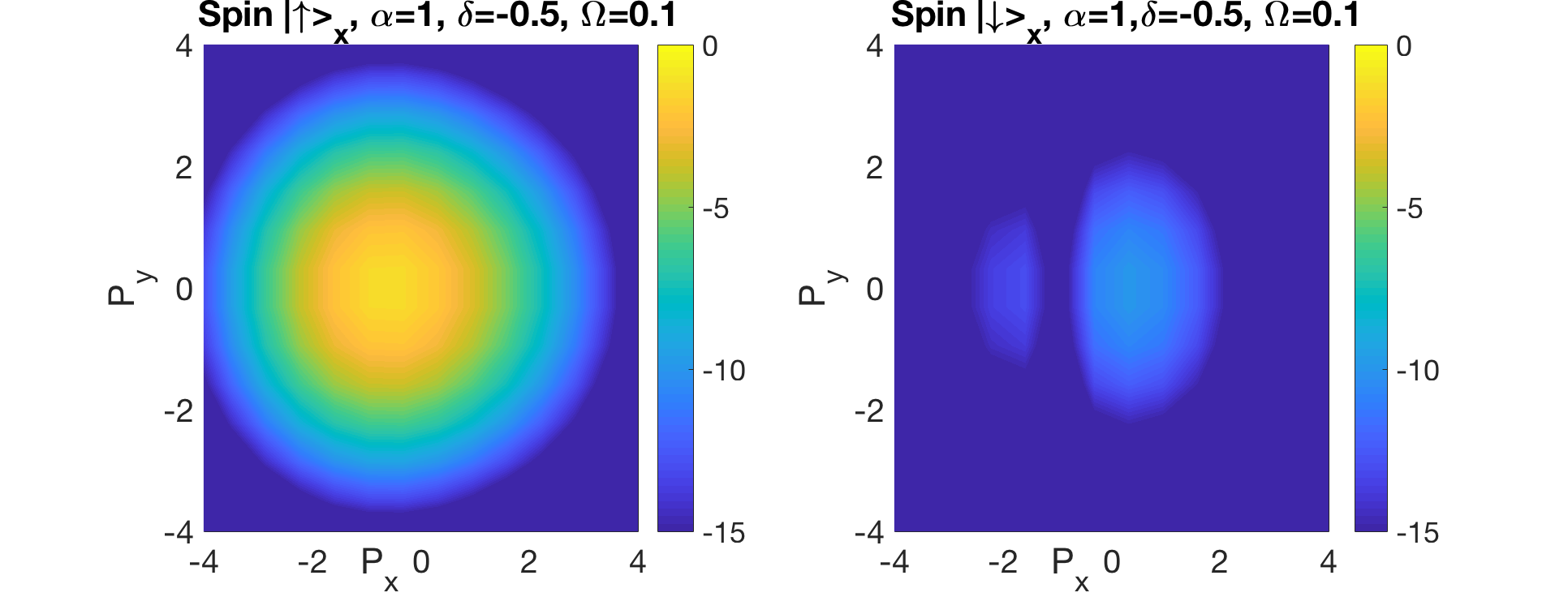}
\caption{(color online) The ground state momentum density when $\delta$ is negative, since there are two local minimum at $k_{x}\neq 0$. Comparing to Fig. \ref{spin_up_phase}, the pairs $|\downarrow_{x},k_{x}>0>, |\uparrow_{x},k_{x}<0>$ are indicated clearly.}
\label{spin_up_phase_p}
\end{figure}

Similar things will happen when $\Omega=0$ and I change $\delta > 0$.
The ground state will prefer the $|\uparrow>_{x}$ components and change the average $k_{x}<0$. If $\Omega$ is not very strong, there will still be some fraction in $|\downarrow_{x},k_{x}>0>$. This result is shown in Fig. \ref{spin_up_phase} and Fig. \ref{spin_up_phase_p}.

From the simulation results, the inter-particle interaction doesn't play an important role here to determine the ground state phase. This is because the SO coupling here is not nearly isotropic. The $k_{x}\sigma_{x}$ term doesn't have a  strong degeneracy here, thus there is no amplifying of the scattering term in the Hamiltonian.

\begin{figure}
\includegraphics[width=0.5\textwidth]{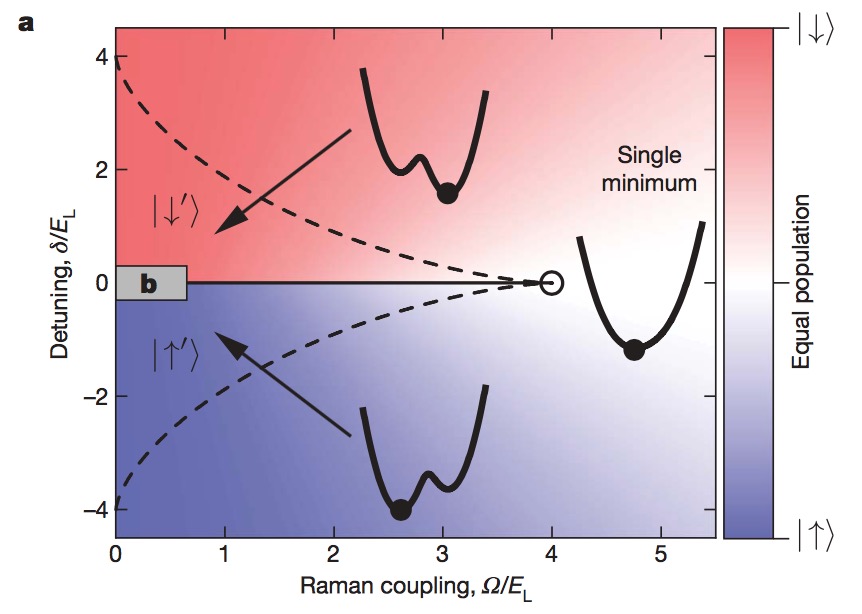}
\caption{(color online) The ground state phase diagram calculated in \cite{spielman}.}
\label{total phase}
\end{figure}

By changing different $\delta$ and $\Omega$, I can control the dispersion minimum. I can realize the results of the simulations by varying the Raman laser coupling $\Omega$ and $z$-direction magnetic field $\delta$ adiabatically in experiment. To detect whether the state is mixed or at different phase, the time of flight method is applied. Through letting the condensates evolve freely after some time, ground states has $k_{x}\neq 0$ at the double minimum phase will split. The total phase diagram in Fig. \ref{total phase} is calculated in \cite{spielman}.

To sum up, I first introduced the definition of spin-orbital coupling in this paper. Then, I discussed the SW and PW phases in a 2D Rashba SO coupled spin-1/2 BEC by controlling the sign of the scattering term \cite{SOtheory} in the first part. The ground state in the PW phase will show the domain wall formed by vortices. This reveals the fact that nearly isotropic SO coupling can amplify the inter-particle interaction. The next topic I discussed is the realization in an experiment by using the Raman coupling lasers in a $^{87}Rb$ spin-1 BEC\cite{spielman}. By taking advantage of this technique, I can study the effect of an equal contribution of Rashba and Dresselhaus coupling. By controlling the Raman coupling strength and magnetic field, the ground state can be tuned adiabatically from a single spin component to double components, and from nonzero momentum to zero momentum. The basic method to do the analysis and numerical simulation is also included also to help understand the results.

\newpage

\clearpage
\bibliography{mybib}

\end{document}